\documentclass[12pt]{article}
\usepackage{latexsym}
\begin{document}
\input epsf
\def\be{\begin{equation}}
\def\bea{\begin{eqnarray}}
\def\ee{\end{equation}}
\def\eea{\end{eqnarray}}
\def\d{\partial}
\def\la{\lambda}
\def\eps{\epsilon}
\def\a{{\cal A}}
\begin{flushright}
OHSTPY-HEP-T-01-018\\
hep-th/0107113
\end{flushright}
\vspace{20mm}
\begin{center}
{\LARGE The  slowly rotating near extremal D1-D5 system as a `hot tube'}
\\
\vspace{20mm}
{\bf  Oleg Lunin  and  Samir D. Mathur \\}
\vspace{4mm}
Department of Physics,\\ The Ohio State University,\\ Columbus, OH 43210, USA\\
\vspace{4mm}
\end{center}
\vspace{10mm}
\begin{abstract}

The geometry of the D1-D5 system with a small angular momentum j has  a
long throat ending in a
conical defect. We solve the  scalar wave equation for low energy
quanta in this geometry. The
quantum is found to reflect off the end of the throat, and stay
trapped in the throat for a long time.  The length of the throat for j=1/2
equals $n_1n_5 R$, the  length of the effective string in the CFT; we also find
that at this distance the incident wave  becomes nonlinear. Filling
the throat with several quanta gives a `hot tube' which has  emission
properties similar to
those of the near extremal black hole.

\end{abstract}
\newpage

\section{Introduction.}
\renewcommand{\theequation}{1.\arabic{equation}}
\setcounter{equation}{0}

String theory has had considerable success in explaining the quantum
properties of black holes. We
do not however have a clear understanding of how the information in
infalling matter
transfers itself to the emerging Hawking radiation, so we do not have
a resolution of the `information paradox'.  It is thus interesting to
probe the physics of
systems that are close to the threshold of  black hole formation,
with a view towards uncovering the
mechanism of information transfer.

The D1-D5 system has been central to our understanding of black holes through
string theory. The D1-D5-momentum bound state yields a microscopic count of
states that agrees exactly with the Bekenstein entropy for extremal \cite{sv}
and near-extremal \cite{cm} black holes. This system also emits, through a
unitary process,  low energy radiation that agrees exactly in spin dependence
and radiation rate with the Hawking radiation expected from the hole \cite{dm}.
In the limit where we take the momentum charge to be much less than the
other two charges one finds that the above agreement persists at higher orders
in the energy of the infalling quantum, and thus emission from the microscopic
system reproduces the grey-body factors of  Hawking radiation
\cite{ms}.

In this paper we look at the D1-D5 system with a small amount of
angular momentum.  The geometry
in the absence of rotation exhibits an  $S^3$  that asymptotes to a  nonzero
size as the radial coordinate $r$ tends to zero. The distance to
$r=0$ is infinite. We will call this
region where the size of the $S^3$ approximates its asymptotic value
as the `throat' of the geometry.
It was shown in
\cite{mm} that the presence of rotation causes the throat  to
truncate at some distance in a conical
defect. For small values of rotation the throat is long. We will
refer to this truncated throat as a
`tube'.

With this system we perform the following investigations:

\bigskip

(a)\quad We consider the wave equation for a massless minimally
coupled scalar in such a geometry.
It is known that the radial equation separates from the angular
equation. Performing this separation
      we  solve the radial equation for long wavelength modes by joining
the solution  in the large $r$ region to the solution in the small $r$
region. It turns out that the wave
reflects naturally off the end of the tube,  and thus we do not have
to choose between different
possible boundary conditions at $r=0$.

Long wavelength modes incident from infinity have  a small
probability for entering the throat, and
the wave reflected from the end of the throat has a similar small
probability for leaving the throat
and escaping back to infinity. We compute the `time delay' from the
wavefunction to find how long
the quantum stays in the `tube'.  We find time delays corresponding
to traveling  $1,2, 3, \dots$
times down the tube  before emerging. These time delays agree with
the travel time estimated from the
path of a null geodesic bouncing off the ends of the tube.

\bigskip

(b) \quad Let $R$ be the radius of the direction which is common to
the D1 and D5 branes, and $n_1,
n_5$ be the number of D1 and D5 branes.  It is known that the length
scale $n_1n_5R$ appears in the
physics of the D1-D5 black hole. Thermodynamically, it is the
wavelength of the last quantum
emitted in a thermal approximation, and microscopically  it is the
length of the
`effective string' formed in the D1-D5 bound state.  In solving the
wave equation mentioned above
we find that this length appears as a feature of the  geometry, in
several different ways, as follows.

Looking at a class of ground states of the D1-D5 system we observe
that the minimal value of angular
momentum for at least this class of states is $j=1/2$ rather than
$j=0$. Thus we look at the geometry
for this minimal value of rotation. We find that in this case the
time needed for a quantum to bounce
off the end of the tube is $\sim Rn_1n_5$.

Next, we consider  the geometry as having no rotation, but
throw in a quantum that carries $j\sim 1/2$. We ask if the
back-reaction of this quantum will cause
horizon formation (due to its energy)  or  truncate the throat in a
conical defect (due to its angular
momentum). We find that the former happens if the wavelength is
smaller than $\sim Rn_1n_5$,
while the latter happens if the wavelength exceeds $\sim Rn_1n_5$.

We then look at the waveform for a spinless quantum propagating in
the geometry with no rotation.
We observe that the amplitude of the wave increases as $\sim
r^{-1/2}$ as $r\rightarrow 0$. We
estimate the point where the wave amplitude becomes large enough to
turn on nonlinear effects.
This point turns out to be a distance $\sim R n_1n_5$  along the throat.
      \bigskip

(c)\quad Since quanta  have a small probability to exit from the
tube, we can trap them for a long
time in the tube and  make a  thermal bath. We refer to this  as a
`hot tube'. We investigate the
properties of such a thermal bath, and observe that it radiates in a
manner similar to a near
extremal black hole.

We conclude the paper with some speculations about the quantum nature
of the geometry deep
down the throat, and implications for the information problem of black holes.

\bigskip

The plan of this paper is as follows. Sections 2, 3  and 4 study the wave
equation in the rotating geometry.
Section 5 discusses the length scale
$Rn_1n_5$.  In section 6 we look at the thermal properties of the hot
tube.  Section 7 examines more general possibilities like `branching
throats'. Section
8 is a discussion where we comment on the information paradox.

\section{Wave equation for the scalar field.}
\renewcommand{\theequation}{2.\arabic{equation}}
\setcounter{equation}{0}

We begin with the metric for the rotating D1--D5 system \cite{cvetic, mm}:
\bea\label{MaldToCompare}
d{s}^2&=&-\frac{1}{h}(d{t}^2-d{ y}^2)+
hf\left(d\theta^2+\frac{d{r}^2}{{r}^2+a^2}\right)
-\frac{2a\sqrt{Q_1 Q_5}}{hf}\left(\cos^2\theta d{ y}d\psi+
\sin^2\theta d{ t}d\phi\right)\nonumber\\
&+&h\left[
\left({ r}^2+\frac{a^2Q_1Q_5\cos^2\theta}{h^2f^2}\right)
\cos^2\theta d\psi^2+
\left({ r}^2+a^2-\frac{a^2Q_1Q_5\sin^2\theta}{h^2f^2}\right)
\sin^2\theta d\phi^2\right],\nonumber \\
\eea
where
\be\label{defFHProp}
f={ r}^2+a^2\cos^2\theta,\qquad
h=\left[\left(1+\frac{Q_1}{f}\right)\left(1+\frac{Q_5}{f}\right)\right]^{1/2}
\ee
The parameter $a$ can be written in terms of the dimensionless variable
$\gamma=\frac{2J}{n_1n_5}$:
\be
a=\frac{\sqrt{Q_1Q_5}}{R}\gamma
\ee
Here $J$ is the angular momentum in units of $\hbar$; thus it is an integer or
half integer. The maximum value of $J$ is ${1\over 2} n_1n_5$, so the maximum
value of $\gamma$ is unity.

       Let us consider a minimally coupled scalar in this
background. The wave  equation for such a scalar reads:
\be\label{GenScalEqn}
\Box\Phi+M^2\Phi=0.
\ee
where:
\be
\Box\Phi=\frac{1}{\sqrt{-g}}\d_\mu
\left(\sqrt{-g}g^{\mu\nu}\d_\nu\Phi\right)
\ee

We can naturally  decompose the
wave operator into three parts:
\be\label{SplitLapl}
\Box\Phi\equiv \Box_{r,\theta}\Phi+\Box_{t,\phi}\Phi+
\Box_{y,\psi}\Phi,
\ee
where each component contains only derivatives with respect to variables
listed in the subscript. For example,
\bea
\Box_{t,\phi}\Phi&\equiv& {1\over
\sqrt{-g}}\left[\d_t\left(\sqrt{-g}g^{tt}\d_t\Phi\right)+
\d_t\left(\sqrt{-g}g^{t\phi}\d_\phi\Phi\right)+
\d_\phi\left(\sqrt{-g}g^{t\phi}\d_t\Phi\right)\right.\nonumber \\
&&\left.+
\d_\phi\left(\sqrt{-g}g^{\phi\phi}\d_\phi\Phi\right)\right]
\eea
The metric (\ref{MaldToCompare}) has four Killing vectors corresponding to
translation along $t$, $\phi$, $\psi$ and $y$ coordinates, so we can look
for the solution  in the following form:
\be
\Phi(t,r,\theta,\phi,\psi,y)=\exp(-i\tilde\omega t+im\phi+in\psi+i\tilde\la y)
{\tilde\Phi}(r,\theta).
\ee

Let us evaluate the three parts of (\ref{SplitLapl}) separately:
\bea
\Box_{r,\theta}\Phi&=&\frac{1}{hf}
\left\{\frac{1}{r}\d_r\left(r(r^2+a^2)\d_r\Phi\right)+
\frac{1}{\sin 2\theta}\d_\theta\left(\sin 2\theta\d_\theta\Phi\right)
\right\}\\
\Box_{t,\phi}\Phi&=&\frac{1}{hf}
\left\{\tilde\omega^2R^2\left[\frac{r^2}{R^2}+\frac{Q_1+Q_5}{R^2}+
\frac{1}{r^2+a^2}\frac{Q_1Q_5}{R^2}
\right]
-\sqrt{Q_1Q_5}
\frac{2m\tilde\omega a}{r^2+a^2}+\frac{m^2a^2}{r^2+a^2}\right.\nonumber\\
&&\left.+\tilde\omega^2 a^2\cos^2\theta
-\frac{m^2}{\sin^2\theta}\right\}\Phi\\
\Box_{y,\psi}\Phi&=&-\frac{1}{hf}
\left\{\tilde\la^2R^2\left[\frac{r^2}{R^2}+\frac{Q_1+Q_5}{R^2}+
\frac{1}{r^2}\frac{Q_1Q_5}{R^2}
\right]+\sqrt{Q_1Q_5}\frac{2n\tilde\la a}{r^2}+
\frac{n^2a^2}{r^2}\right.\nonumber\\
&&\left.+\tilde\la^2 a^2\cos^2\theta+
\frac{n^2}{\cos^2\theta}\right\}\Phi
\eea
We can see that for general value of the mass $M$ the variables $r$
and $\theta$
remain coupled in (\ref{GenScalEqn}), but for massless particles we have a
separation. If we consider
\be
\tilde\Phi(r,\theta)=H(r)\Theta(\theta),
\ee
then the Klein--Gordon equation (\ref{GenScalEqn}) for the massless case:
\bea
\Box\Phi=0
\eea
can be rewritten as a system:
\bea\label{ProperRad}
&&\frac{1}{r}\frac{d}{dr}\left(r(r^2+a^2)\frac{dH}{dr}\right)+
\left\{\tilde\omega^2R^2\left[\frac{r^2}{R^2}+\frac{Q_1+Q_5}{R^2}+
\frac{1}{r^2+a^2}\frac{Q_1Q_5}{R^2}
\right]\right.\nonumber\\
&&\qquad-\sqrt{Q_1Q_5}
\frac{2m\tilde\omega a}{r^2+a^2}+\frac{m^2a^2}{r^2+a^2}
-\sqrt{Q_1Q_5}\frac{2n\tilde\la a}{r^2}-
\frac{n^2a^2}{r^2}\nonumber\\
&&\qquad\left.-\tilde\la^2R^2\left[\frac{r^2}{R^2}+\frac{Q_1+Q_5}{R^2}+
\frac{1}{r^2}\frac{Q_1Q_5}{R^2}
\right]\right\}H-\Lambda H=0
\eea
\bea\label{ProperAngul}
\frac{1}{\sin 2\theta}\frac{d}{d\theta}\left(\sin 2\theta
\frac{d\Theta}{d\theta}\right)
+\left\{
a^2(\tilde\omega^2-\tilde\la^2)\cos^2\theta-
\frac{m^2}{\sin^2\theta}
-\frac{n^2}{\cos^2\theta}\right\}\Theta=-\Lambda\Theta
\eea
The angular equation reduces to the Laplacian on   $S^3$ if
$a\tilde\omega,a\tilde\lambda\rightarrow 0$, which allows us to determine the
approximate eigenvalues:
\be\label{EigenLaAppr}
\Lambda=l(l+2)+O((a\tilde\omega)^2)+O((a\tilde\lambda)^2)
\label{rone}
\ee
where $l$ is a nonnegative integer.  We work out the value of
$\Lambda$ to order $a^2$ in
Appendix A, but we note here that we will need only the leading term
(\ref{rone}).  The
existence of a separation of variables for such a wave equation  was shown in
\cite{larsentwo}.

Let us simplify the radial equation. We introduce a new coordinate
\be
x=\frac{r^2 R^2}{Q_1Q_5}
\ee
and dimensionless parameters $\omega=R\tilde\omega$, $\la=R\tilde\la$.
Then the radial equation (\ref{ProperRad}) becomes
\bea\label{RadEqn}
&&4\frac{d}{dx}\left(x(x+\gamma^2)\frac{dH}{dx}\right)+
\left\{(\omega^2-\la^2)\left[\frac{Q_1Q_5}{R^4}x+\frac{Q_1+Q_5}{R^2}\right]
\right.\nonumber\\
&&\qquad\left.+\frac{(\omega-m\gamma)^2}{x+\gamma^2}
-\frac{(\la+n\gamma)^2}{x}
\right\}H-\Lambda H=0
\eea
      From now on we will restrict ourselves to the case $\lambda=0$;
i.e., we will
not allow any momentum along the compact direction $y$.

\section{Solution of the wave equation.}
\renewcommand{\theequation}{3.\arabic{equation}}
\setcounter{equation}{0}

\subsection{ The matching technique}

While we can not solve the complete radial
equation (\ref{RadEqn}), it is  possible to solve it in two
asymptotic regions and
match solutions in an intermediate region for low frequencies $\omega$.  This
technique has been used several times in the past to study absorption
into black
holes
\cite{absorption, wadia, dm, ms,KlMath}.

\subsection{Solution in the outer region.}

In the outer region ($x$ large), the equation (\ref{RadEqn}) becomes:
\be
4x^2H''+8xH'+\omega^2
\left\{\frac{Q_1Q_5}{R^4}x+\frac{Q_1+Q_5}{R^2}\right\}H-\Lambda H=0
\ee
The general solution of this equation can be written in terms of Bessel's
functions:
\be\label{OuterHyperAc}
H_{out}(x)=\frac{1}{\sqrt{x}}\left[C_1
J_\nu\left(\sqrt{\frac{Q_1Q_5\omega^2x }{R^4}}\right)+
C_2
J_{-\nu}\left(\sqrt{\frac{Q_1Q_5\omega^2x }{R^4}}\right)\right],
\ee
where
\be\label{DefNu}
\nu=\left(1+\Lambda-\omega^2\frac{Q_1+Q_5}{R^2}\right)^{1/2}\equiv
l+1+\eps
\ee
and
\be
\eps\approx -\frac{\omega^2}{l+1}\frac{Q_1+Q_5}{2R^2}
\label{eone}
\ee

We have chosen to define the parameter $\epsilon$ here for the
following reason.
We have dropped a term proportional to ${\omega^2\over x} H$ in writing the
wave equation in the outer region. However when using the matching
technique, we will use this solution in a region where $x$ is small (though
${\omega\over \sqrt x}\ll 1$).  The value of
$\epsilon$ given in (\ref{eone}) arises from retaining the term proportional to
$\omega^2 H$, which is smaller than the term we drop in the matching region.
In fact none of the terms containing $\omega^2$ are significant in the matching
region, and we could have dropped all of them. Keeping one of the terms does
not improve the accuracy of the matching, but allows us to use the basis of two
Bessel functions $J_\nu$ and $J_{-\nu}$ since $\nu$ is shifted away from an
integer by the amount $\epsilon$. (We find this trick more convenient than
using a basis of $J$ and $N$ functions.) Thus $\epsilon$ is a
regulator which allows
us to use a nearly degenerate basis of functions, and we will see
explicitly that
$\epsilon$ cancels at the end of our computations.

       Using the series expansion
for the Bessel's functions:
\be\label{ExpandBess}
J_{\mu}(z)=\left(\frac{z}{2}\right)^\mu\sum_{n=0}^\infty
\frac{1}{\Gamma(\mu+n+1)n!}
\left[-\left(\frac{z}{2}\right)^2\right]^n,
\ee
we can express $J_{-\nu}(z)$ in the following form:
\bea\label{ExpandF01}
J_{-\nu}(z)&=&\left(\frac{z}{2}\right)^{-\nu}\left[\sum_{n=0}^{l}
\frac{(-z^2/4)^n}{\Gamma(-l-\eps+n)n!}+
\sum_{n=l+1}^{\infty}\frac{(-z^2/4)^n}{\Gamma(-l-\eps+n)n!}\right]\\
&=&\left(\frac{z}{2}\right)^{-l-1}\left[-(-1)^l\eps l!\left(1+O(z^2)\right)+
\frac{1}{(l+1)!}\left(-\frac{z^2}{4}\right)^{l+1}\left(1+O(z^2)+O(\eps)\right)
\right]\nonumber
\eea
Here we have used the fact that
$\Gamma(-l-\eps)=-\left[(-1)^l \eps l!\right]^{-1}$.

We will make an expansion of (\ref{OuterHyperAc}) in the region where
\be
\frac{z^2}{4}\equiv \frac{Q_1Q_5\omega^2x }{4R^4}\ll 1.
\ee
We also will use the fact that $\eps\ll 1$. However we will not make any
assumptions about a magnitude of the ratio
\be
\frac{z^{2l+2}}{\eps}
\ee
and we will keep both leading terms in (\ref{ExpandF01}). For the Bessel
function $J_\nu(z)$ we will keep only the leading term corresponding to
$n=0$ in the expansion (\ref{ExpandBess}).
Then for
(\ref{OuterHyperAc}) we get:
\bea\label{AsympOut}
H_{out}(x)\approx\frac{1}{\sqrt{x}}
\left(\frac{Q_1Q_5\omega^2x }{4R^4}\right)^{-\frac{l+1}{2}}
\left[-(-1)^l C_2\eps l!+
\frac{C_1-(-1)^l C_2}{(l+1)!}\left(\frac{Q_1Q_5\omega^2x }{4R^4}\right)^{l+1}
\right].
\eea

\subsection{Solution in the inner region.}

Let us now look at the inner region where
\be
x\ll \frac{(Q_1+Q_5)R^2}{Q_1Q_5}
\ee
Then (\ref{RadEqn}) becomes:
\bea\label{RadEqnPr}
&&4\frac{d}{dx}\left(x(x+\gamma^2)\frac{dH}{dx}\right)+
\left\{\omega^2\left[\frac{Q_1+Q_5}{R^2}\right]
+\frac{(\omega-m\gamma)^2}{x+\gamma^2}
-\frac{(n\gamma)^2}{x}
\right\}H-\Lambda H=0
\eea
We write $H(x)$ as:
\be
H_{in}(x)=x^\alpha(\gamma^2+x)^\beta G(x),
\ee
where
\be
\alpha=\frac{n}{2},\qquad
\beta=\frac{\omega-\gamma m}{2\gamma}
\ee
Then the radial equation becomes
\bea\label{radialEqnGamma}
&&4x(x+\gamma^2)G''+4\left[2x(\alpha+\beta+1)+\gamma^2(1+2\alpha)\right]
G'\nonumber\\
&&+\left\{4(\alpha+\beta)(\alpha+\beta+1)+
\omega^2\frac{Q_1+Q_5}{R^2}-\Lambda
\right\}G=0
\eea
This is the hypergeometric equation.  There is a regular and a
singular solution
at $x=0$. We choose the regular solution for the following reason. We have a
conical defect singularity only where $f=r^2+a^2\cos^2\theta=0$, which
requires $r=0, \theta={\pi\over 2}$. But for $r=0$ and $\theta\ne {\pi\over 2}$
we have no singularity. While we cannot assume that the solution is
regular at a
geometric singularity like a conical defect, we need the solution to
be regular at
       $r=0,\theta\ne {\pi\over 2}$. Since the solution factorizes in the
$r$, $\theta$
variables, we find that the function of $x\sim r^2$ must be regular at $x=0$ to
ensure  such regularity. We will comment on this issue further at the
end of this
section.

The solution of (\ref{radialEqnGamma}) regular at
$x=0$ reads
\bea\label{SolveHyperG}
G(x)&=&F\left(p,q;
1 + 2\alpha;-\frac{x}{\gamma^2}\right),\\
p&=&\frac{1}{2}+\alpha+\beta +
\frac{1}{2}\sqrt{1+\Lambda-\omega^2\frac{Q_1+Q_5}{R^2}}\\
q&=&\frac{1}{2}+\alpha+\beta -
\frac{1}{2}\sqrt{1+\Lambda-\omega^2\frac{Q_1+Q_5}{R^2}}
\eea
Note that we have already encountered the square root appearing in this
expression in (\ref{DefNu}) where it was called $\nu$. We had
       introduced the parameter $\eps=\nu-l-1$. and noted that it served as a
regulator allowing us to use a nearly degenerate basis of Bessel functions. We
have a similar situation here, and again the precise value of the square root
in
the above relations is not significant since it would be
significantly affected
by terms that have been dropped. Again we note that all terms
proportional to $\omega^2$ can be dropped in the region of matching.  We write
\bea
p=\frac{1}{2}+\alpha+\beta+
\frac{1}{2}(l+1+\eps')\nonumber\\
q=\frac{1}{2}+\alpha+\beta-
\frac{1}{2}(l+1+\eps')
\eea
where we have chosen a different symbol $\epsilon'$ to denote the regulator
allowing us to use a nearly degenerate set of functions in the inner
region.  We
will verify that at the end of the calculations the regulators
$\epsilon, \epsilon'$
cancel independently in the results.
We  also define
\be\label{NuPrime}
\nu'=l+1+\eps'
\ee

To summarize, in the region $x\ll \frac{(Q_1+Q_5)R^2}{Q_1Q_5}$, the solution
of the radial equation
(\ref{RadEqnPr}) is
\be\label{SolveHyper}
H_{in}(x)=x^\alpha(\gamma^2+x)^\beta F\left(p,q;
1 + 2\alpha;-\frac{x}{\gamma^2}\right)
\ee

We will need the asymptotic form  of this expression for  large
values of $x$.
To get this asymptotics it is convenient to rewrite (\ref{SolveHyperG}) using
an identity relating hypergeometric functions:
\bea\label{ExactG}
G(x)&=&\frac{\Gamma(1 + 2\alpha)\Gamma(-{\nu'})}{
\Gamma(\frac{1}{2}+\alpha+\beta-\frac{{\nu'}}{2})
\Gamma(\frac{1}{2}+\alpha-\beta-\frac{{\nu'}}{2})}
\left(\frac{x}{\gamma^2}\right)^{-p}
F\left(p,p-2\alpha;{\nu'}+1;-\frac{\gamma^2}{x}\right)\nonumber\\
&+&\frac{\Gamma(1 + 2\alpha)\Gamma({\nu'})}{
\Gamma(\frac{1}{2}+\alpha+\beta+\frac{{\nu'}}{2})
\Gamma(\frac{1}{2}+\alpha-\beta+\frac{{\nu'}}{2})}
\left(\frac{x}{\gamma^2}\right)^{-q}
F\left(q,q-2\alpha;-{\nu'}+1;-\frac{\gamma^2}{x}\right)\nonumber \\
\eea
We will now use the fact that $p\approx q\approx \beta\gg 1$, since we
take $\gamma$ very small so that we have a very long
throat. Let us look at  the following series:
\be
F(p,p-2\alpha;c;-z)=\sum_{n=0}^\infty \frac{\Gamma(p+n)}{\Gamma(p)}
\frac{\Gamma(p-2\alpha+n)}{\Gamma(p-2\alpha)}\frac{\Gamma(c)}{\Gamma(c+n)}
\frac{(-z)^n}{n!}
\ee
We are interested in a case where $p\rightarrow\infty$, while $xp^2$ remains
bounded.   Thus in the above expression we can replace
\be
\frac{\Gamma(p+n)}{\Gamma(p)}\qquad\mbox{by}\qquad p^n
\ee
and we get
\be
F(p,p-2\alpha;c;-z)\approx \sum_{n=0}^\infty \frac{\Gamma(c)}{\Gamma(c+n)}
\frac{(-zp^2)^n}{n!}=(p\sqrt{z})^{-(c-1)}\Gamma(c)J_{c-1}(2p\sqrt{z})
\ee
Then (\ref{ExactG}) can be approximated by
\bea
G(x)&\approx&\frac{\Gamma(1 + 2\alpha)\Gamma(-{\nu'})\Gamma({\nu'}){\nu'}}{
\Gamma(\frac{1}{2}+\alpha+\beta+\frac{{\nu'}}{2})
\Gamma(\frac{1}{2}+\alpha-\beta+\frac{{\nu'}}{2})}
\left(\frac{x}{\gamma^2}\right)^{-\frac{n-m+1}{2}-\frac{\omega}{2\gamma}}
p^{-{\nu'}}\\
&\times&\left[
\frac{\Gamma(\frac{1}{2}+\alpha+\beta+\frac{{\nu'}}{2})
\Gamma(\frac{1}{2}+\alpha-\beta+\frac{{\nu'}}{2})}
{\Gamma(\frac{1}{2}+\alpha+\beta-\frac{{\nu'}}{2})
\Gamma(\frac{1}{2}+\alpha-\beta-\frac{{\nu'}}{2})}
J_{\nu'}\left(\frac{2\gamma p}{\sqrt{x}}\right)
-(pq)^{\nu'} J_{-{\nu'}}\left(\frac{2\gamma q}{\sqrt{x}}\right)\right]
\nonumber
\eea
Using the relations
$\gamma^2 p^2\approx \gamma^2 q^2\approx \omega^2/4$, we can rewrite the last
expression as
\be
G(x)\approx D_1J_{\nu'}\left(\frac{\omega}{\sqrt{x}}\right)+
D_2J_{-{\nu'}}\left(\frac{\omega}{\sqrt{x}}\right),
\ee
and the radial wavefunction as
\be\label{InnnerH}
H_{in}(x)\approx \frac{1}{\sqrt{x}}\left[
D_1J_{\nu'}\left(\frac{\omega}{\sqrt{x}}\right)+
D_2J_{-{\nu'}}\left(\frac{\omega}{\sqrt{x}}\right)\right],
\ee
where
\be\label{RatD1D2}
\frac{D_1}{D_2}=-(pq)^{-{\nu'}}
\frac{\Gamma(\frac{1}{2}+\alpha+\beta+\frac{{\nu'}}{2})
\Gamma(\frac{1}{2}+\alpha-\beta+\frac{{\nu'}}{2})}
{\Gamma(\frac{1}{2}+\alpha+\beta-\frac{{\nu'}}{2})
\Gamma(\frac{1}{2}+\alpha-\beta-\frac{{\nu'}}{2})}
\ee
Note that if we were solving the problem without rotation (i.e. $\gamma=0$),
then (\ref{InnnerH}) would give the solution for the entire inner region (which
in that case becomes an infinite throat), although we would not be able to find
       any relation between $D_1$ and $D_2$ since there would be  no natural
boundary condition from the end of the throat $x=0$.

It is convenient to rewrite the RHS of (\ref{RatD1D2}) in terms of gamma
functions with large positive arguments. Using the expression
\be
\Gamma(x)=\frac{\pi}{\sin\pi x}\frac{1}{\Gamma(1-x)},
\ee
we get:
\be\label{RatD1D2p}
\frac{D_1}{D_2}=-(pq)^{-{\nu'}}
\frac{\Gamma(\frac{1}{2}+\alpha+\beta+\frac{{\nu'}}{2})
\Gamma(\frac{1}{2}+\beta-\alpha+\frac{{\nu'}}{2})}
{\Gamma(\frac{1}{2}+\alpha+\beta-\frac{{\nu'}}{2})
\Gamma(\frac{1}{2}+\beta-\alpha-\frac{{\nu'}}{2})}~
\frac{\sin\left[\pi(\alpha-\beta+(1-\nu')/2)\right]}
{\sin\left[\pi(\alpha-\beta+(1+\nu')/2)\right]}
\ee
To extract the leading order in $1/\beta$ we use an approximate relation
\be
\frac{\Gamma(\frac{1}{2}+\alpha+\beta+\frac{{\nu'}}{2})}
{\Gamma(\frac{1}{2}+\alpha+\beta-\frac{{\nu'}}{2})}\approx \beta^{\nu'},
\ee
then using the expression (\ref{NuPrime}) for $\nu'$, we can rewrite the ratio
(\ref{RatD1D2p}) as
\be
\frac{D_1}{D_2}=(-1)^l
\frac{\sin\left[\pi\left(\beta-\alpha+l/2+\eps'/2\right)\right]}
{\sin\left[\pi\left(\beta-\alpha+l/2-\eps'/2\right)\right]}
\ee
which can be rewritten as
\be\label{DRatio}
\frac{D_1-D_2(-1)^l}{D_2(-1)^l}=\pi\eps'\frac{
\cos\left[\pi\left(\beta-\alpha+l/2\right)\right]}
{\sin\left[\pi\left(\beta-\alpha+l/2\right)\right]}
\ee

In the next subsection we will match the solutions between the  inner and outer
regions.  To do this we will need the behavior of (\ref{InnnerH}) for $x\gg
\omega^2$ (which can be derived in the same way as the asymptotic expression
(\ref{AsympOut}))
\bea\label{AsympIn}
H_{in}(x)\approx\frac{1}{\sqrt{x}}
\left(\frac{\omega^2}{4x}\right)^{-\frac{l+1}{2}}
\left[-(-1)^lD_2\eps' l!+
\frac{D_1-(-1)^lD_2}{(l+1)!}\left(\frac{\omega^2}{4x}\right)^{l+1}
\right]
\eea

\subsection{Matching solutions in the intermediate region.}

Let us compare the solutions (\ref{AsympOut}) and (\ref{AsympIn}). Requiring
that
\be
H_{out}(x)\approx H_{in}(x)
\ee
we get
\bea\label{MatchCoeff}
\frac{C_2}{C_1-(-1)^l C_2}&=&\frac{D_1-D_2(-1)^l}{D_2}\frac{1}{\eps\eps'}
\left(\frac{Q_1Q_5\omega^4 }{16R^4}\right)^{l+1}
\left[\frac{1}{(l+1)!l!}\right]^2\\
\label{MatchCoeff1}
C_2&=&\frac{D_2-(-1)^lD_1}{\eps l!(l+1)!}
\left(\frac{Q_1Q_5\omega^4}{16R^4}\right)^{\frac{l+1}{2}}
\eea

\subsection{The nature of the boundary condition at $r=0$}

One might at first think that since the geometry has a conical defect
at $r=0$ one will need
to choose one out of several possible boundary conditions at this
point before the wave
problem is well defined. But we have seen that the boundary condition
is automatically
determined from the nature of the equation. At an algebraic level,
this happened because
the singularity is only at $r=0, \theta=\pi/2$ rather than for all
points with $r=0$. This
might still have implied that we need a boundary condition at $r=0,
\theta=\pi/2$, but the
equation factorizes between  $r$ and $\theta$, so that demanding
regularity of the solution
at $r=0, \theta\ne \pi/2$ determines the behavior also at $r=0, \theta=\pi/2$.

It would however be good to understand more physically how the wave
turned back from
$r=0$. To do this let us analyze the wave equation (\ref{RadEqn}) for
$x\rightarrow 0$. We replace an expression like $x+\gamma^2$ by $\gamma^2$.
Then the equation
becomes
\be
4\gamma^2\frac{d}{dx}\left(x\frac{dH}{dx}\right)+
\left\{(\omega^2-\la^2)\frac{Q_1+Q_5}{R^2}
+\frac{(\omega-m\gamma)^2}{\gamma^2}-\Lambda
-\frac{(\la+n\gamma)^2}{x}
\right\}H=0
\ee
Writing $z=-\log x$, we can write the above equation as
\be
-{d^2H\over dz^2}+V(x)H=0
\ee
\be
V(x)={(\la+n\gamma)^2\over 4\gamma^2}-{e^{-z}\over
4\gamma^2}[(\omega^2-\la^2)\frac{Q_1+Q_5}{R^2}
+\frac{(\omega-m\gamma)^2}{\gamma^2}-\Lambda]\equiv  P-Qe^{-z}
\ee

Thus we have a Schroedinger equation with the above potential,  with
total energy $E=0$.
$x\rightarrow 0$ corresponds to $z\rightarrow \infty$. Note that
$P\ge 0$. If $P>0$, the
wavefunction with $E=0$ will automatically reflect back from the
potential barrier back
towards larger $r$. If $P=0$, then we can solve the equation by
writing $H=1+c_1 e^{-z} +
c_2 e^{-2z}+\dots$ and obtain a convergent expansion. The
wavefunction is real, and no
flux is carried out to $r=0$ so that we see again that we get a
reflection from $r=0$.

This situation is to be contrasted with the equation we get when
there is no angular
momentum and  the throat is infinitely long. Then for small $r$ the
equation takes the
form
\be
{d\over dx}(x^2 {dH\over dx})+{\kappa\over x}H=0
\ee
with $\kappa={1\over 4}[\omega^2-\la^2]$. Writing $y=1/x$
we get a Schroedinger equation of the form
\be
-{d^2H\over dy^2}-{\kappa\over y}H=0
\ee
As we have seen, the solution can be expressed in terms of Bessel functions,
and a wave
incident from infinity becomes
oscillatory at $y\rightarrow\infty$, carrying flux in towards $r=0$.

Roughly speaking we may say that if a region of the geometry  pinches
off too sharply
(as at the conical defect) then the wave reflects back from the walls
of the throat near the
pinch, while if it narrows more slowly then the wave continues along
the throat, possibly
gaining in amplitude, but in any case carrying flux in towards $r=0$.
Note however that if
we depart from the supergravity approximation and require stringy
effects to set in at
string length, then it is possible to modify the above found
behavior of the wavefunction
in a region of radius string length around $r=0$.  In that case the
incident wave could be
absorbed at the conical defect and possibly return with some knowledge of the
quantum state at the defect. (A similar reflection from the singularity for 
a different rotating system was studied in \cite{GibHerd}).

\section{Phase shift in the far outer region.}
\label{SectShift}
\renewcommand{\theequation}{4.\arabic{equation}}
\setcounter{equation}{0}

We have solved the low energy wave equation in the metric
(\ref{MaldToCompare}). From this solution we wish to extract the
nature of scattering in
this geometry; more precisely, we wish to use the method of phase
shifts to compute a time
delay and thus see how long the particle stays in the throat. Before
we analyze the phase
shifts it is helpful to understand the physics we expect, and to do
this we first solve a toy
model that produces similar phase shifts.

A quantum incident from infinity enters the throat with a small
probability; the rest of the
wave reflects back to infinity from the  start of the throat. The
wave which does enter the
throat travels to its end, reflects back, but then has a small
probability to emerge from the
throat out to infinity. The rest of this trapped wave travels back
down the throat again. The
emerging wave thus has a large proportion with time delay zero --
this is the part that
never entered the throat. A small part will have time delay $2L$,
where $L$ is the length
of the throat in the coordinate where the speed is unity. But there
will be additional parts
with time delays $4L, 6L, \dots$.  It is easy to see that the parts
with time delay will have
amplitude squared proportional $\a^2$, where $\a$ is the probability
to enter or leave the
throat.

We model this situation by a 1+1 dimensional Klein-Gordon equation in
the following way.
We put a delta function  potential barrier  at $x=0$ which reflects
most of the wave incident
from
$x>0$, but allows a small part to enter the region $x<0$. This latter
region is the analogue of
the throat. At
$x=-L$ we put a reflecting wall, modeling the end of the throat.

\subsection{Toy problem.}
\label{SubsToy}

Let us consider a toy system which is governed by an analog of the two
dimensional Klein--Gordon equation:
\be\label{KleinG}
\left(\frac{\d^2}{\d t^2}-\frac{\d^2}{\d x^2}\right)\Phi(x,t)+V(x)
\Phi(x,t)=0
\ee
with potential $V(x)$ given by
\bea
V(x)&=&c\delta(x),\qquad x>-L\\
V(x)&=&\infty,\qquad x<-L
\eea
We will look for the solution in the form: $\Phi(x,t)=e^{-ikt}\Psi(x)$. Then
(\ref{KleinG}) can be rewritten as
\be
-\Psi''(x)+V(x)\Psi(x)=k^2\Psi(x)
\ee
We can solve this equation in three different regions:
\bea
x>0:\qquad \Psi(x)&=&Ae^{ikx}+Be^{-ikx},\\
0>x>-L:\qquad \Psi(x)&=&Ce^{ikx}+De^{-ikx},\\
x<-L:\qquad \Psi(x)&=&0.
\eea
and use boundary condition at $x=0$ to write $A$ and $B$ in terms of $C$ and
$D$:
\bea
A=C-\frac{ic}{2k}(C+D),\qquad B=D+\frac{ic}{2k}(C+D).
\eea
In particular we will need the ratio
\be\label{ToyRatio}
R\equiv\frac{A}{B}=
\left(1+\frac{2ik}{c}\frac{C}{C+D}\right)
\left(-1+\frac{2ik}{c}\frac{D}{C+D}\right)^{-1}.
\ee
The boundary condition $\Psi=0$ at $x=-L$ gives:
\be
C=-e^{2ikL}D
\ee
Then we get for the ratio (\ref{ToyRatio}):
\be\label{ToyRatioFin}
R=
\left(1-\frac{2ik}{c}\frac{e^{2ik L}}{1-e^{2ik L}}\right)
\left(-1+\frac{2ik}{c}\frac{1}{1-e^{2ik L}}\right)^{-1}
\ee

   Let us rewrite the ratio (\ref{ToyRatioFin}) in the following
form:
\bea
R&=&-\left(\frac{1}{1-\frac{2ik}{c}}-e^{2ikL}
\frac{1+\frac{2ik}{c}}{1-\frac{2ik}{c}}\right)
\left(1-\frac{e^{2ikL}}{1-\frac{2ik}{c}}\right)^{-1}\nonumber\\
&=&-\left(\frac{1}{1-\frac{2i k}{c}}-e^{2ikL}
\frac{1+\frac{2i k}{c}}{1-\frac{2i k}{c}}\right)
\sum_{n=0}^\infty \left(\frac{e^{2ikL}}{1-\frac{2i k}{c}}\right)^n
\nonumber\\
&=&-\frac{1}{1-\frac{2i k}{c}}+
\left(\frac{2 k}{c}\right)^2\sum_{n=1}^{\infty}
\frac{e^{2iknL}}{\left(1-\frac{2i k}{c}\right)^{n+1}}
\eea
This gives the expression for
the wavefunction $\Phi_k$
corresponding to the given frequency $k$:
\be\label{PhiKWave}
\Phi_k(x,t)=B\left[e^{-ik t-ik x}+e^{-ik t+ik x}
\left(
-\frac{1}{1-\frac{2i k}{c}}+
\left(\frac{2 k}{c}\right)^2\sum_{n=1}^{\infty}
\frac{e^{2iknL}}{\left(1-\frac{2i k}{c}\right)^{n+1}}
\right)\right]
\ee
If we look at a wave packet, for example,
\be
\Phi(x,t)=\int dk \Phi_k(x,t)e^{-k^2/\la},
\ee
we will see the incoming wave with a peak at $x=-t$ as well as various
outgoing waves. The leading outgoing wave has a peak at $x=t$, and it
corresponds to scattering back from the delta function potential. But there are
also ``secondary waves'' with $x=t-2Ln$, which correspond to  scattering
from the wall at $x=-L$. At the leading order in $\frac{2k}{c}$
the probability of going beyond $x=0$ and coming
back is given by
\be\label{ResP1}
P_1=\left(\frac{2 k}{c}\right)^4.
\ee
We can also find the probability $S_1$ of  an incident particle to
proceed beyond
$x=0$ by comparing
$B$  with $D$. The calculation is done in the Appendix \ref{AppB} and
the result is
$S_1=\sqrt{P_1}$,  as we could anticipate on physical grounds.

\subsection{Phase shift for the D1--D5 system.}
\label{SubsD1D5Shift}

Let us go back to the solution in the outer region (\ref{OuterHyperAc})
\be
H_{out}(x)=\frac{1}{\sqrt{x}}\left(C_1
J_\nu\left(\sqrt{\frac{Q_1Q_5\omega^2x }{R^4}}\right)+
C_2
J_{-\nu}\left(\sqrt{\frac{Q_1Q_5\omega^2x }{R^4}}\right)\right).
\ee
At very large $x$ we can use the asymptotics of the Bessel's functions:
\be
J_{\nu}(z)=\sqrt{\frac{2}{\pi z}}\cos(z-\frac{\pi\nu}{2}-\frac{\pi}{4})+
O(z^{-1}),
\ee
then we get:
\bea\label{FFarout1}
H_{out}&=&\frac{1}{x^{3/4}}
\left(\frac{4R^4}{Q_1Q_5\omega^2\pi^2}\right)^{1/4}
\left[C_1\cos\left(\omega y-\frac{\pi\nu}{2}-\frac{\pi}{4}\right)
+C_2\cos\left(\omega y+\frac{\pi\nu}{2}-\frac{\pi}{4}\right)\right]
\nonumber\\
&+&
O(x^{-3/4-1/2})
\eea
Here we have introduced a natural variable in the far outer region:
\be
y=\sqrt{\frac{Q_1Q_5x}{R^4}}
\ee

We now need to compare the phases of the ingoing and outgoing waves.
Ignoring  the overall $y$ dependence of the wavefunction we have
\bea
H_{out}&\sim& e^{i\omega y-i\pi\frac{l}{2}-\frac{3\pi i}{4}}
\left(C_1e^{-i\pi\eps/2}-C_2(-1)^le^{i\pi\eps/2}\right)\nonumber\\
&&~~~+~~e^{-i\omega y+i\pi \frac{l}{2}+\frac{3\pi i}{4}}
\left(C_1e^{i\pi\eps/2}-C_2(-1)^le^{-i\pi\eps/2}\right)
\eea
Since we are looking for a long time delay we can ignore the finite
shifts in $y$
included in the exponentials in the above expression. Following what
we did in the
toy problem   we consider the ratio:
\be
R=\frac{C_1e^{-i\pi\eps/2}-C_2(-1)^le^{i\pi\eps/2}}
{C_1e^{i\pi\eps/2}-C_2(-1)^le^{-i\pi\eps/2}}
\ee
Note that for real $C_1$ and $C_2$ we get $|R|=1$, in accordance with
the fact that
there is no loss of flux down the throat.
We rewrite
$R$ in the following form:
\bea
R&=&e^{-i\pi\eps}-(1-e^{-2i\pi\eps})\frac{C_2(-1)^l}
{C_1-C_2(-1)^l e^{-i\pi\eps}}\nonumber\\
&\approx& e^{-i\pi\eps}-2i\pi\eps
\frac{D_1-D_2(-1)^l}{D_2(-1)^l}\frac{1}{\eps\eps'}
\left(\frac{Q_1Q_5\omega^4 }{16R^4}\right)^{l+1}
\left[\frac{1}{(l+1)!l!}\right]^2 \nonumber \\
&\approx& e^{-i\pi\eps}-2i\pi^2
\frac{\cos\left[\pi\left(\beta-\alpha+l/2\right)\right]}
{\sin\left[\pi\left(\beta-\alpha+l/2\right)\right]}
\left(\frac{Q_1Q_5\omega^4 }{16R^4}\right)^{l+1}
\left[\frac{1}{(l+1)!l!}\right]^2
\eea
where in the second step we have used the expression
(\ref{MatchCoeff}) to relate
$C$ and
$D$, and in the last step we have used (\ref{DRatio}).

To find the different phase shifts contained in the scattered wave we
write the above
expression in a form similar to  (\ref{PhiKWave})
\be\label{ApprR}
R\approx e^{-i\pi\eps}-2\pi^2\frac{1+
e^{2\pi i\left(\beta-\alpha+l/2\right)}}
{1-e^{2\pi i\left(\beta-\alpha+l/2\right)}}
\left(\frac{Q_1Q_5\omega^4 }{16R^4}\right)^{l+1}
\left[\frac{1}{(l+1)!l!}\right]^2
\ee
and make a formal expansion of the term $1-e^{2\pi
i\left(\beta-\alpha+l/2\right)}$
in the denominator:
\bea
R&\approx& \left[e^{-i\pi\eps}-
2\pi^2\left(\frac{Q_1Q_5\omega^4 }{16R^4}\right)^{l+1}
\left[\frac{1}{(l+1)!l!}\right]^2\right]\nonumber\\
&-&4\pi^2
\left(\frac{Q_1Q_5\omega^4 }{16R^4}\right)^{l+1}
\left[\frac{1}{(l+1)!l!}\right]^2\sum_{n=1}^{\infty}
e^{2\pi in\left(\beta-\alpha+l/2\right)}
\eea
  From the phases of the terms in the infinite sum we read off the time
delay between
the emerging wave--packets:
\be
\Delta t=2\pi\frac{\d}{\d\tilde\omega}(\beta-\alpha)=
\pi\frac{R}{\gamma}=\pi\frac{\sqrt{Q_1Q_5}}{a}.
\ee
The coefficient of the scattered waves yields  the probability of going
into the throat and coming back
\be
P_2=\left[4\pi^2\left(\frac{Q_1Q_5\omega^4 }{16R^4}\right)^{l+1}
\left[\frac{1}{(l+1)!l!}\right]^2\right]^2
\ee

We can also evaluate the probability $S_2$ for going inside the throat. As in
the case of the toy model we get $P_2=S_2^2$. The details of the calculation
are presented in  Appendix \ref{AppB}.  The result should agree with the known
probability of absorption into the near extremal D1-D5 geometry
\cite{ms}, where
the particle enters the throat and then proceeds inwards without
returning. As a
cross check on our calculations we verify that we indeed obtain the correct
probability.

\section{Properties of the `hot tube'.}
\renewcommand{\theequation}{5.\arabic{equation}}
\setcounter{equation}{0}

\subsection{Some scales in the D1-D5 system}

Let us first recall the relations between the microscopic D1-D5 system
(with no angular momentum) and the
geometry that it produces. We set the momentum charge $n_p$ of the system to
zero, but allow a small nonextremality characterized by the parameter $r_0$.
We follow for the most part the notations of \cite{hms}, \cite{ms}.

The 10-d Einstein metric and dilaton are
\bea
ds^2&=&(1+{Q_1\over r^2})^{-3/4}(1+{Q_5\over
r^2})^{-1/4}\left[-(1-{r_0^2\over r^2})dt^2+dx_5^2+(1+{Q_1\over
r^2})dx_idx_i\right]\nonumber\\
&+&(1+{Q_1\over r^2})^{1/4}(1+{Q_5\over
r^2})^{3/4}\left[(1-{r_0^2\over r^2})dr^2+r^2d\Omega_3^2\right]
\label{qone}
\eea
\be
e^{-2\phi}=(1+{Q_1\over r^2})^{-1}(1+{Q_5\over
r^2})
\ee
We write
\bea
Q_1&\equiv& r_1^2~=~r_0^2\sinh^2(2\alpha_1)\nonumber \\
Q_5&\equiv& r_5^2~=~r_0^2\sinh^2(2\alpha_5)
\eea
We work in units where $\alpha'=1$.
The volume of the $T^4$  spanned by the coordinates $x_i$ is $(2\pi)^4 V$, and
the length of the circle
$x_5$ is
$2\pi R$. The 10-d Newton's constant is $G_{10}=8\pi^6g^2$.  The extremal
configuration is given by $r_0\rightarrow 0, \alpha_1\rightarrow\infty,
\alpha_5\rightarrow\infty$, while holding $Q_1, Q_5$ fixed. We will work in the
near extremal limit $r_0\ll r_1, r_5$.

The number of D1 branes and D5 branes that
produce the above solution are respectively
\be
n_1={Vr_0^2\over 2g}\sinh (2\alpha_1)\approx {VQ_1\over g}
\ee
\be
n_5={r_0^2\over 2g}\sinh (2\alpha_5)\approx{Q_5\over g}
\ee

The mass of the solution is
\be
M={RVr_0^2\over 2g^2}(\cosh(2\alpha_1)+\cosh(2\alpha_5)+1)
\ee
where the last term in the bracket arises from a term $\cosh(2\alpha_p)$ when
we set $\alpha_p=0$. (The parameter $\alpha_p$ gives the momentum
charge through $n_p=(R^2Vr_0^2/2 g^2)\sinh (2\alpha_p)$.)   For small
nonextemality we find that the mass above extremality is essentially given by
the last term in the above relation
\cite{memission}
\be
\delta M\equiv M-M_{extremal}\approx {RVr_0^2\over 2g^2}
\label{qthree}
\ee

For our case of no momentum charge the left and right temperatures $T_L,
T_R$ of the system equal the Hawking temperature $T_H$
\be
T_L=T_R=T_H={r_0\over 2\pi r_1r_5}.
\label{qfour}
\ee

\subsection{The length scale $Rn_1n_5$}

The thermodynamic properties of the  D1-D5 system are well described by an
`effective string' which is `multiply wound' around $x_5=y$.  We are interested
in the ground state which corresponds to one single long string; this string
has  a total effective length
$L_{eff}=n_1n_5 L$,  where $L=2\pi R$ is the length of the $x_5$ circle. It was
argued in
\cite{dmmw} that the low energy excitations of a multiwound string are
harmonic vibrations that carry left and right momentum in units of
$2\pi/L_{eff}$, but that the net momentum of the system must still be
an integer
multiple of
$2\pi/L$.  The lowest excitation thus has one left and
one right moving mode, with net momentum zero and a total energy
\be
\delta M={4\pi\over L_{eff}}={2\over n_1n_5 R}
\label{qfive}
\ee
It was argued in \cite{msuss} that the above energy scale is a physically
relevant scale emerging from the thermodynamics of the D1-D5 geometry: if
the nonextremal energy is not much larger than (\ref{qfive}) then the
temperature suffers a fractional change of order unity when a single typical
quantum is emitted, and  the emission process ceases to be described by
naive thermodynamics \cite{preskill}.  We may verify this fact as follows.
We start from
the extremal system ($r_0=0$), throw in a quantum of energy $\tilde\omega$ and
ask that the temperature increase from zero to order $\tilde\omega$. We have
\be
\tilde\omega=\delta M\sim {RVr_0^2\over g^2}
\ee
while
\be T_H\sim {r_0\over r_1r_5}.
\ee
Requiring $\tilde\omega\sim T_H$ gives the desired result
\be
\tilde\omega\sim {1\over R n_1n_5}.
\ee

For future use, let us define
\be
      \tilde\omega_0\equiv {1\over R n_1n_5}.
\ee

\subsubsection{The length $n_1n_5R$ from the length of the throat}

The unexcited D1-D5  system can be in one of several Ramond ground states, all
of which have the same energy.  These Ramond (R) ground states can be
obtained by spectral flow of chiral primary states from the Neveu-Schwarz
(NS) sector.  The chiral primary states can be placed in a finite number of
classes,  based on which cohomology element we pick from the compact
4-manifold (which can be $T^4$ or $K3$). Let us consider the description of
states at the orbifold point of the D1-D5 system. The  states  involve  twist
operators $\sigma_n$ of order $n$.  A basic family of states, studied
in \cite{lmSUSY},  has
dimension and SU(2) charge $h=j_3={n-1\over 2}=j$. After we spectral flow such
a state to the R sector we get $h={c\over 24}, j_3={n-1\over 2}-{c\over 12}$.
Since $c=6N, N=n_1n_5$, we get $h={N\over 4}$ and
\be j_3={n-1-N\over 2}
\label{wone}
\ee
      The R sector states
arrange themselves into multiplets of SU(2) with (\ref{wone}) as the lowest
value of the $j_3$ charge.
But since $1\le n\le N$, we see that at least for this family of
states the smallest
value of SU(2) charge is $j=1/2$ and not $j=0$.

Note that this SU(2) charge is just the angular momentum of the state. But if
the angular momentum is given by $\gamma\ne 0$ then we know that the
throat of the corresponding geometry is not infinite but in fact truncated at
some distance with the formation of a conical defect. Let us set the angular
momentum to be
$j=1/2$, so that
\be
\gamma={j\over j_{max}}={{1\over 2}\over {1\over 2} n_1 n_5}={1\over
n_1n_5}\equiv
\gamma_0
\label{wtwo}
\ee

Let us now ask how long the throat is for this value of $\gamma$. We imagine a
massless particle thrown into the throat, and ask how long it takes
to reach the
end. Note that $\gamma=\gamma_0$ gives a very small angular momentum to
the system, so we may use the geometry without angular momentum to
compute the time of flight of the particle, if we set $x\approx
\gamma_0^2$ as the
end of the throat. This corresponds to $r\approx \gamma_0\sqrt{Q_1Q_5}/R$. We
set
$Q_1\sim Q_5$. The start of the throat may be taken as
$r\sim (Q_1Q_5)^{1/4}$.  The time of flight is then given by setting $ds^2=0$,
which gives
\be\label{EstimT}
t=\int_{r_{min}}^{r_{max}} dr {g_{rr}^{1/2}\over -g_{tt}^{1/2}}\approx
\int_{\gamma_0\sqrt{Q_1Q_5}/R}^{(Q_1Q_5)^{1/4}} \sqrt{Q_1Q_5} ~{dr \over
r^2}\approx n_1n_5 R
\label{wthree}
\ee

Thus we see that the length of the tube in the geometry with minimum
angular momentum gives the wavelength scale associated to the point where
thermodynamic behavior of the near extremal D1-D5 system breaks down.

\subsubsection{Horizon formation}

Consider the geometry of the extremal D1-D5 system with no rotation
($\gamma=0$).  Let us
imagine that we throw in a massless quantum with energy $\omega$,
with a spin $j=1/2$.
As the quantum travels down the throat it can reach a point where
its mass will cause a horizon to
form. But because the quantum carries nonzero $j$, the geometry
outside the location of
the quantum will exhibit this value of angular momentum and so the
throat can at some point end
in a conical defect. Let us ask if the horizon will form first or the
conical defect will form first. If the
conical defect forms before the horizon, then we may speculate that
the particle will bounce back
along the throat rather than disappear behind a horizon, though we
cannot prove this rigorously
since we are not explicitly solving for the backreaction created by
the quantum.

     To  locate the point of horizon formation  we need to find the
value of $r_0$ when the nonextremal mass given to the D1-D5 system is
$\tilde\omega$. Setting
\be
\tilde\omega = \delta M \sim {RVr_0^2\over g^2}
\ee
we get
\be
r_0^2\sim {g^2 \tilde\omega\over RV}
\ee
which corresponds to
\be
x={R^2\over Q_1Q_5}r_0^2\sim {1\over
(n_1n_5)^2}=\gamma_0^2{\tilde\omega\over \tilde\omega_0}
\ee

Thus we observe that if $\tilde\omega=\tilde\omega_0=1/Rn_1n_5$ then the
horizon
forms at $x\sim
\gamma_0^2$, which is the same point where the conical defect would
terminate the tube because of
the angular momentum carried by the quantum. If $\tilde\omega>\tilde\omega_0$
(and the particle still has
$j=1/2$) then the horizon forms before the rotation terminates the
tube. For energies
$\tilde\omega<\tilde\omega_0$ we may speculate that the quantum returns back
without horizon formation,
which would agree with the fact that in the microscopic picture of
the `effective string' energies
lower than $1/Rn_1n_5$ cannot be absorbed.

\subsubsection{Nonlinearity of the wave equation}

Let us now arrive at the length scale $Rn_1n_5$ in yet another way.
Consider the geometry of the unexcited D1-D5 system with $\gamma=0$; the
geometry has an infinite throat. Let us throw in one quantum of a scalar field.

The infall of the quantum down the throat can be described, at least for some
time, by the linear wave equation studied in the last section. But as the wave
moves down the throat and acquires an oscillatory character, we see that its
amplitude also increases. This can be seen from the behavior of the
solution in terms of Bessel's functions valid for the nonrotating
geometry in the
throat region. The solution behaves as
\be
\phi \sim D_1 {1\over \sqrt{x}} J_\nu ({\omega \over \sqrt x})+D_2 {1\over
\sqrt{x}} J_{-\nu} ({\omega \over \sqrt x})
\ee
But for large argument the Bessel's functions itself behaves as
\be
J_\nu(z)\approx \sqrt{2\over \pi z} \cos(z-{\pi\over 2}\nu -{\pi\over 4})
\ee
so that we see that the amplitude of the wave grows in the throat as
\be
x^{-1/4}\sim r^{-1/2}
\ee
The start of the throat is at $r\sim (Q_1Q_5)^{1/4}$. Let us follow the quantum
down the throat for the time $Rn_1n_5$, which brings us to $r\sim
\gamma_0\sqrt{Q_1Q_5}/R$. This implies an amplification factor
\be
\mu=[{\gamma_0(Q_1Q_5)^{1/4}\over R}]^{-1/2}
\label{wfive}
\ee

To see whether the wave has at this point become essentially nonlinear, we
must know the initial strength of the wave. To estimate this we write
the action
for the scalar as
\be
S={1\over 16\pi G_N} \int d^{10} x {1\over 2} \partial\phi \partial\phi
\ee
and we obtain the Hamiltonian from this action.
The energy of a quantum is $\tilde\omega$. To estimate the value of
$\phi$ for such a
quantum at the start of the throat, we need to know the volume $d^9x$
occupied by the wavepacket. We may confine the wavepacket to a longitudinal
distance of order $\tilde\omega^{-1}$, while the transverse area is
made up of $V$
(the volume of the $T^4$), $R$ (the length of $y$), and the area of
$S^3$ which at
the start of the throat is $\sim (Q_1Q_5)^{3/4}$. Thus we get
\be
{1\over G_N}\tilde\omega^{-1}VR(Q_1Q_5)^{3/4}(\tilde\omega^2\phi^2)\sim
\tilde\omega
\ee
Note that $\tilde\omega$ cancels out, and we get (using
$G_N\sim g^2$)
\be
\phi\sim [{g^2\over RV(Q_1Q_5)^{3/4}}]^{1/2}
\ee
Multiplying by the amplification factor (\ref{wfive}) we find that the value of
$\phi$ at the point $x\sim \gamma_0^2$ is
\be
\phi\sim 1
\ee
so that the linear wave equation can become invalid after the
quantum travels a distance
$Rn_1n_5$ down the throat.

\section{Making the D1-D5 system into a `hot tube'}
\renewcommand{\theequation}{6.\arabic{equation}}
\setcounter{equation}{0}

      We have seen that a quantum traveling down the throat of the D1-D5 system
reflects off the end if $\gamma\ne 0$. The quantum then travels back  towards
the start of the throat. Our calculations were done for low energies,
so that the
wavelength of the incoming quantum was  much larger than the length scale
$(Q_1Q_5)^{1/4}$ describing the scale of variation of the geometry at
the start of
the throat.  When such a low
energy quantum is thrown in from infinity, then there is a small probability
that it enters the throat. Let us restrict our attention  to s-waves
($l=0$). The
absorption probability  is
$\a \sim \tilde\omega^4Q_1Q_5\ll 1$.  But this also implies that when the
quantum
reflects from the end of the throat and travels back to larger $r$
then it has the
same probability $\a$ to escape out of the throat to infinity. Thus the wave
reflects back (with a probability close to unity) to small $r$ again,
and we see that
the quantum is trapped in the throat for a long time. We can fill up the throat
with several quanta, and in particular imagine that the configuration
is placed in
a thermal state. (Whether an actual initial configuration of quanta reaches
thermal equilibrium before effusing out depends on  the strength of
interactions and the length of the tube.) We call such a thermal
configuration a
`hot tube'. In this section we study some of the properties of these hot tubes.
(In \cite{msfive} the near extremal 5-brane was studied. The throat
in this case was
also filled with radiation, but it ended in a horizon since there was
no rotation.)

\subsection{Some preliminaries}

      Let us imagine that we
fill up the tube with radiation that has a temperature $T$. The energy of a
typical quantum will be $\tilde \omega \sim T$. We measure all
energies in terms
of the time coordinate at infinity, and recall that the frequency in
this coordinate
was called $\tilde \omega\equiv \omega/R$.

The length of the tube is determined by the value of the parameter $\gamma$;
the smaller the value of $\gamma$, the longer the tube. We do not
want to take too long a
tube, since otherwise a horizon will form at some point in the tube,
and we will
have a black hole which swallows quanta rather than an endpoint that reflects
them. Let the total nonextremal energy be $E_T$. Then the position of
the horizon
would be
\be
r_0^2\sim {g^2 E_T\over RV}, ~~\rightarrow~~x\sim {g^2 RE_T\over VQ_1Q_5}
\ee
(Of course if the radiation is distributed through the tube rather than
concentrated at the end then not all the energy $E_T$ will contribute
to horizon
formation. But it turns out that reducing the value of $E_T$  to take
into account
this effect does not change the above estimates.) The tube ends at $x\sim
\gamma^2$, so that we need
\be
\gamma^2 > {g^2 R E_T\over V Q_1Q_5}.
\ee
Note that putting $E_T=(R n_1n_5)^{-1}$ gives $\gamma\sim 1/n_1n_5$.

We also do not want the tube to be too short since we want the wavelength of
the typical quantum to fit into the tube. The solution of the wave
equation in the
throat was given in terms of Bessel's functions with argument ${\omega\over
\sqrt{x}}$, and these functions become oscillatory in the throat only for
${\omega\over
\sqrt{x}} >1$. Since the throat ends at $x\sim \gamma^2$, we get the
requirement
\be
\gamma <\omega=R\tilde\omega  \sim RT
\ee

\subsection{The relation between $E_T$ and $T$}

We still need to find the relation between the energy added to the
tube $E_T$ and
the temperature $T$ which the tube attains as a consequence; i.e., we need to
know the heat capacity of the tube. To do this it is helpful to
recall the expression
for the time of flight for a massless particle in the throat:
\be
t=\sqrt{Q_1Q_5}\int  {dr\over r^2}
\ee
Let us write
\be
y={\sqrt{Q_1Q_5}\over r}
\ee
Then the classical trajectory for a massless particle satisfies
$dt=|dy|$ and the
waveform behaves as
$e^{-i\tilde \omega (t\pm y)}$. We have thus a regular one dimensional thermal
system.  The minimum value of $y$ is $\sim (Q_1Q_5)^{1/4}$ at the start of the
throat, but we can set this to zero for our estimates. The end of the
tube is at
$y\sim {R\over \gamma}$. The number of wavelengths of the typical quantum
that fit in this interval of $y$ is
\be
n\sim {R/\gamma\over \tilde\omega^{-1}}={R\tilde \omega\over \gamma}
\ee
The total energy of the gas of quanta is then
\be
E_T\sim \tilde \omega n\sim {R\tilde \omega^2\over \gamma}\sim {R
\over \gamma} T^2
\ee

[If we consider instead the system with no rotation but a horizon at some
$r=r_0$ then the nonextremal energy is related to the temperature by
\be
E_{BH}\sim {R\over \gamma_0}T^2, ~~~~\gamma_0={1\over n_1n_5}
\ee
It is curious that if we naively set $\gamma=\gamma_0$ (its minimum value)
then we reproduce with the hot tube  the order of the heat capacity
of the black
hole. But if we choose $\gamma$ to be so small then the backreaction of the
thermal quanta will lead to horizon formation in the tube and invalidate the
tube analysis. It is possible that there is some deeper reason which
tells us that
we can ignore the  backreaction and recover the  properties of the
black hole.]

\subsection{Radiation from the hot tube}

Quanta can slowly leak out of the throat of this hot tube just as
radiation escapes
to infinity from the near extremal black hole. The essential difference is of
course that the quanta in the throat are generated by the horizon in the black
hole case, while they have been added by hand for the hot tube. Thus the
temperature of the hot tube is not determined by its geometry, while it is
determined by geometry for a black hole.

We are looking at the $l=0$ harmonics (where most of the radiation occurs). The
probability for a  quantum incident from infinity to enter  the throat is $\a$.
After the quantum enters the throat, it travels to a very good approximation as
a free particle with the waveform $e^{-i\tilde\omega(t-y)}$, before reflecting
from the end of the tube. Thus the probability for it to exit the tube is
determined by local physics near the start of the throat ($r\sim
(Q_1Q_5)^{1/4}$) and unrelated to the geometry at the end of the tube. We have
therefore a free gas of massless quanta trapped in the tube, which gives a
1-dimensional thermal system.  each time a quantum reaches back to the start
of the throat it has a probability $\a$ of escaping to infinity. We
now estimate the
radiation rate from the tube.

Let each typical quantum have a wavelength $\lambda$. Let the effective length
along the tube (measured in the coordinate $y$ where $dt=|dy|$) be
$L=n\lambda$. A given quantum reaches the start of the tube $\sim 1/n\lambda
$ times per unit of the time $t$. There are $\sim n$ quanta of this typical
wavelength in the thermal distribution so $1/\lambda\sim \tilde\omega$
quanta try to escape the tube per unit time. Since the escape
probability is $\a$,
the number of quanta exiting per unit time is
\be
\Gamma\sim \tilde\omega A
\ee

Let us compare this rate to the emission rate from a near extremal black hole
which has a temperature $T\sim \tilde \omega$. The number of quanta emitted
per unit time is
\be
\Gamma_{BH} = {\sigma\over e^{\tilde\omega/T}-1} {d^4k\over (2\pi)^4}
\ee
But the absorption cross section $\sigma$ is given by $\sigma={4\pi\over
\tilde\omega^3}\a$ where $\a$ is the same probability as the one appearing in
the calculation with the hot tube.  Setting the Boltzmann factor to
be order unity
for a typical quantum (we did the same for the hot tube) we get
\be
\Gamma_{BH}\sim {\a\over \tilde\omega^3}\tilde \omega^4 \sim \tilde\omega \a
\ee
which is of the same order as emission from the hot tube. This is not
surprising,
since the emission from the throat depends on the temperature in the throat,
and we have compared systems where this temperature is the same. We have
seen above though that the heat capacity is in general different in
the two cases.

\section{Throat geometry and states of the D1-D5 CFT}
\renewcommand{\theequation}{7.\arabic{equation}}
\setcounter{equation}{0}

We have seen above that the geometry of the throat  reflects  the
properties of the D1-D5 bound state  in the following way. The D1-D5
system can be
modeled by an effective string that is wrapped $n_1n_5$ times around the
circle of radius $R$.
If all the strands of this string are joined together to make one
long strand, then the lowest
energy mode is $\tilde \omega\sim 1/R n_1n_5$. We have seen that if
we go a distance
$Rn_1n_5$ down the throat of the corresponding geometry then special
effects occur: a
spin $j=1/2$ would end the throat in a conical defect, and any scalar
wave would tend to
become nonlinear. But the D1-D5 system can be in a host of different
states -- the spin
may not be $j=1/2$ or all the strands of the effective string may not
be joined into  a single
string. In this
section we briefly discuss how the throat geometry might reflect
these possibilities.

\bigskip

{\it Spins along $T^4$:}

\bigskip

Let the strands of the effective string be joined up into one single
long string. This string
has a multiplet of ground states. A subset of these ground states
have spin $j$ along the
$S^3$ directions of order unity, and we have seen that this spin
causes the throat to end at
a distance $\sim n_1n_5 R$. There are other members of the multiplet
which have no
spin along $S^3$, but have a spin along the directions of the compact
space $T^4$.  Let us
consider such states.

We note that the $S^3$ gives the $R$ symmetry group $SO(4)$, but in the
$T^4$ the local symmetry is another $SO(4)$, which we call $SO(4)_I$.
The compactification breaks rotational symmetry in the $T^4$, but for
our rough analysis
we ignore this fact. These two $SO(4)$'s are known to play very
similar roles in the system,
and so we speculate that a spin along the $T^4$ would cause the
throat to truncate in the
same way (and at the same approximate distance) as the spin $j=1/2$
along the $S^3$.

\bigskip

{\it Branching throats:}

\bigskip

Now consider a state of the CFT where the $n_1n_5$ strands of the
effective string are not
all joined together, but are in fact joined to make $m$ different
strings, each of total length
${n_1n_5\over m} R$.  In this case the lowest energy absorbed by the
state would be
$m\tilde\omega_0={m\over n_1n_5 R}$.  We now ask if the throat geometry
could be such so as
to return back (without horizon formation) all modes with energies lower than
$m\tilde\omega_0$.

First consider the case where the spins of each string are along
$S^3$, and further that
these spins are all aligned. Then the state has angular momentum
$j=m/2$, which gives
$\gamma=m\gamma_0$. Recalling the calculation  (\ref{wthree}), and noting
that $r$ at the endpoint is linear in the value of $\gamma$,  we see
that in this case the time
of travel to the end of the throat will be
$\sim {n_1n_5R\over m}$, and with the endpoint of the throat at this
location, quanta with
energy less than $m\tilde\omega_0$ will indeed return without horizon
formation.

Now consider the case where the spins are not aligned, and in fact
add together to give a
total angular momentum $j\sim 0$. The CFT state still suggests that the minimum
frequency that can be absorbed in $m\tilde\omega_0$ (and not $\tilde\omega_0$).
But if we put
$\gamma\sim 0$ or $\gamma\sim 1/2$ in the geometry then we find that
the throat has an
effective length $\sim n_1n_5 R= \tilde\omega_0^{-1}$ or larger. In
view of the fact
that the CFT state was
described by $m$ strings it appears natural to consider that the
throat itself branches at
some distance into
$m$ throats, each with an effective value of $n_1n_5$ given by
${n_1n_5\over m}$. Each
throat also carries $j=1/2$, and the spins of different branches are
not necessarily aligned.
This spin $j=1/2$ causes each branch to terminate after a distance
${n_1n_5 R\over m}$,
and we recover the desired low energy behavior.

Branching throat geometries were also considered in
\cite{strominger} with somewhat
different motivations.
It is possible that quanta with different spins and energies stay 
trapped at different points
along the throat due to the backreaction that they produce, so we 
should not put their energy
and charge together to create the standard black hole geometry.  It 
may also be illuminating to
study absorption by the dual geometry made from winding and momentum 
charges. Here we
may see the string separating into  several strings
\cite{lmRotStr}.

\section{Discussion}
\renewcommand{\theequation}{8.\arabic{equation}}
\setcounter{equation}{0}

The near horizon geometry of the D1-D5 system is locally $AdS_3\times S^3$
\cite{maldacena}.  If the direction $y$ is taken noncompact, then the
space $r>0$ occupies a
Poincare patch of the
$AdS$ spacetime. The geometry smoothly continues to $r<0$ however,
suggesting that
there is at least one more region of spatial infinity\cite{hor}. This appears
paradoxical, since  we
can assemble D1 and D5 branes in flat space  to make the D1-D5
geometry, and it is hard to
see how a new infinite region can emerge just by bringing together
solitons in a space with
only one spatial infinity.

When the direction $y$ is compactified, the geometry is expected to
become singular
at $r=0$ since the points identified under $y\rightarrow y+2\pi R$
become closer and closer
to each other as $r\rightarrow 0$.  Thus we may speculate that in
this quanta thrown
towards $r=0$ return back to the same spatial infinity where they
came from. If a horizon
forms, then we have the usual difficulties of tracking how the
information escapes back to
infinity. But modes that are sufficiently low in energy may return
back without horizon
formation.

The question therefore arises: what is the length scale which governs
when such modes
turn back? One may naively speculate that waves reflect  when the
distance between
identified points $y, y+2\pi R$ becomes smaller than string length or
Planck length. But as
we can see from our discussion this would be too short a return time
scale to agree with the
physics of the D1-D5 microstate. Our results suggest that the throat
effectively
terminates for different reasons, which are related to angular
momentum and nonlinearity
of the wave equation. Our results also suggest that the length of the
throat would equal
the length of the effective string in the D1-D5  CFT, and that a
general description of the
throat geometry might have to include branching of throats.

Finally, we turn to the question of what these results might say for
the problem
of information recovery in the case that a horizon does form. If
throats can branch then
they can also rejoin, and  we imagine a complex web of quantum
fluctuating virtual tubes
that would replace the simple picture of a single infinite throat. In
particular if a tube
rejoins the geometry near but just outside the horizon, then it can
bring information out to
a point where it can
     escape to infinity. There is a cost in action to create a branching
or a joining of tubes, but
near the horizon this cost is low since the proper time for which a
virtual tube lasts is small
due to the redshift factor. We must of course face the usual
difficulty that one can
Lorentz transform to regular variables near the horizon and so
nothing special can happen
there, but such a boost does not make sense if the geometry has a
complex topology of tubes
interconnecting disparate locations. We hope to
return to a study of this issue later.

\newpage
\appendix
\section{Corrections to the eigenvalues of the angular Laplacian.}
\renewcommand{\theequation}{A.\arabic{equation}}
\setcounter{equation}{0}

In this paper we have used an approximate expression for the eigenvalue
$\Lambda$:
\be
\Lambda=l(l+2)
\ee
where $l$ is non--negative integer. This expression becomes exact only for
non--rotating system (i.e. for the metric (\ref{MaldToCompare}) with $a=0$).
In this appendix we will
look at the general equation (\ref{ProperAngul}) for eigenvalue $\Lambda$
and we will
find the
leading order correction to $\Lambda$. For simplicity we will consider only
solutions with $m=n=0$. Then (\ref{ProperAngul}) becomes:
\bea\label{AppEqn1}
\frac{1}{\sin 2\theta}\frac{d}{d\theta}\left(\sin 2\theta
\frac{d\Theta}{d\theta}\right)
+
a^2(\tilde\omega^2-\tilde\la^2)\cos^2\theta\Theta=-\Lambda\Theta
\eea
It is convenient to introduce a new variable $z=\cos 2\theta$. Then the domain
$0\le \theta\le\pi/2$ becomes $-1\le z\le 1$, and equation (\ref{AppEqn1})
becomes:
\be\label{AppEqn3}
4\frac{d}{dz}\left((1-z^2)
\frac{d\Theta}{dz}\right)
+
a^2(\tilde\omega^2-\tilde\la^2)\frac{1+z}{2}\Theta=-\Lambda\Theta
\ee
We are looking for a solution of this equation which is normalizable on the
interval $-1\le z\le 1$.

We will treat $a$--dependent term in (\ref{AppEqn3}) as a small perturbation.
First we note that (\ref{AppEqn3}) can be rewritten as a familiar quantum
mechanical problem
\be\label{QMHam}
(\hat H_0+\hat H_1)|\Theta\rangle=E|\Theta\rangle
\ee
where
\be
\hat H_0\equiv 4\frac{d}{dz}\left((1-z^2)
\frac{d}{dz}\right)
\ee
is the unperturbed Hamiltonian,
\be
\hat H_1\equiv a^2(\tilde\omega^2-\tilde\la^2)\frac{1+z}{2}
\ee
is a perturbation, and
\be
E\equiv-\Lambda.
\ee
Then the unperturbed equation $\hat H_0|\Theta\rangle=E_0|\Theta\rangle$
reads
\be
\frac{d}{dz}\left((1-z^2)
\frac{d\Theta}{dz}\right)+\frac{\Lambda_0}{4}\Theta=0.
\ee
This is the equation for the Legendre polynomials, and the solution
normalizable
on the interval $(-1,1)$ exists only for
\be
\frac{\Lambda_0}{4}=s(s+1),
\ee
where $s$ is a non--negative integer. Note that these eigenvalues correspond to
even angular momentum in (\ref{EigenLaAppr}) ($l=2s$), and to find the
remaining
eigenvalues one should consider the case where the difference $m-n$ is an odd
integer. We will not consider this case here.

The eigenvalue $\Lambda_0=4s(s+1)$ corresponds to the eigenfunction
\be
\Theta_s (z)=P_s(z),
\ee
where $P_s(z)$ is the Legendre polynomial. These polynomials form a complete
system in the space of functions normalizable on the interval $(-1,1)$.

To determine the leading order correction to $\Lambda_0$ we will go back to
the quantum mechanical analogy (\ref{QMHam}) and recall that the lowest order
correction to $E$ is given by
\be
\delta E=\frac{\langle \Theta_0|\hat H_1|\Theta_0\rangle}
{\langle \Theta_0|\Theta_0\rangle},
\ee
where $|\Theta_0\rangle$ is an eigenstate of the unperturbed Hamiltonian
$\hat H_0$. Thus for our problem we get:
\be\label{AppDelz}
\delta \Lambda=-\left(\int_{-1}^1 dz P_s(z)P^*_s(z)\right)^{-1}
\int_{-1}^1 dz a^2(\tilde\omega^2-\tilde\la^2)\frac{1+z}{2}P_s(z)P^*_s(z)
\ee
Let us now use the symmetry properties of the Legendre polynomials:
\bea
P_s(-z)&=&P_s(z)\qquad\mbox{for even }s,\nonumber\\
P_s(-z)&=&-P_s(z)\qquad\mbox{for odd }s.
\eea
In particular, the product $P_s(z)P^*_s(z)$ is an even function of $z$ for
all $s$. Then using the facts that $f(z)=z$ is an odd function of $z$ and
that the ranges of integration in (\ref{AppDelz}) are symmetric under
$z\rightarrow -z$, we get a simple expression for $\delta \Lambda$:
\be
\delta \Lambda=-\left(\int_{-1}^1 dz P_s(z)P^*_s(z)\right)^{-1}
\int_{-1}^1 dz a^2(\tilde\omega^2-\tilde\la^2)\frac{1}{2}P_s(z)P^*_s(z)=
-\frac{a^2}{2}(\tilde\omega^2-\tilde\la^2).
\ee

\section{Evaluation of the absorption probability.}
\label{AppB}
\renewcommand{\theequation}{B.\arabic{equation}}
\setcounter{equation}{0}

In section \ref{SectShift} we have evaluated probabilities of 
reaching and bouncing off
the infinite wall (for the toy system) and off the conical defect at 
the end of the throat. We
called these probabilities $P_1$ and $P_2$.
Here we will evaluate the probability $S_1$ for a quantum incident 
from infinity to pass
through the initial barrier and enter the region $x<0$
  for the toy model and the probability $S_2$ to enter  the throat  in 
the D1-D5 geometry.
On physical grounds we expect
\bea
P_1&=&S_1^2\\
P_2&=&S_2^2
\eea
and we will verify that we indeed obtain these relations. In 
particular $S_2$ will be the
probability for an incident quantum to enter the extremal D1-D5 throat.

We begin with the toy model defined in the subsection \ref{SubsToy}.
Using the relation
\be
B=D\left(1+\frac{ic}{2k}(1-e^{2ikL})\right),
\ee
we find that the wave which looks like
\be
\Phi_k(x,t)=B\left[e^{-ik t-ik x}+\frac{A}{B}e^{-ik t+ik x}\right]
\ee
for positive $x$, in the inner region becomes:
\bea
\Phi_k(x,t)&=&-iB\frac{2k}{c}
\frac{1}{1-\frac{2ik}{c}-e^{2ikL}}
\left[e^{-ik t-ik x}-e^{2ikL}e^{-ik t+ik x}\right]\nonumber\\
&=&-iB\frac{2k}{c}\frac{1}{1-\frac{2ik}{c}}\sum_{n=0}^{\infty}
\left(\frac{e^{2ikL}}{1-\frac{2ik}{c}}\right)^n
\left[e^{-ik t-ik x}-e^{2ikL}e^{-ik t+ik x}\right]
\eea
Thus at the lowest order in $\frac{2k}{c}$ the probability for going
beyond $x=0$ is
\be
S_1=\left(\frac{2k}{c}\right)^2
\ee
Comparing this result with (\ref{ResP1}) we find that $P_1=S_1^2$ as we could
anticipate from the beginning.

Let us now look at absorption by the rotating D1--D5 system. As in the case
of the toy model, we need to evaluate the ratio of coefficients in
front of the left movers in the inner and outer region:
\be\label{OneGoIn}
R'=\frac{D_1e^{i\pi\eps'/2}-D_2(-1)^le^{-i\pi\eps'/2}}
{C_1e^{i\pi\eps/2}-C_2(-1)^le^{-i\pi\eps/2}}
\ee
We will ignore the case of resonance frequencies, then in the leading order in
$\eps$ and $\eps'$ we will replace $e^{\pm i\pi\eps/2}$ and
$e^{\pm i\pi\eps/2}$ by $1$. Then denominator of (\ref{OneGoIn}) becomes
$C_1-C_2(-1)^l$ and after substituting this expression from
(\ref{MatchCoeff}), we get
\be
R'=\left[D_1-D_2(-1)^l\right]\frac{D_1-D_2(-1)^l}{D_2C_2}\frac{1}{\eps\eps'}
\left(\frac{Q_1Q_5\omega^4 }{16R^4}\right)^{l+1}
\left[\frac{1}{(l+1)!l!}\right]^2.
\ee
Taking into account the value of $C_2$ (\ref{MatchCoeff1}), we rewrite this
expression as
\be
R'=-\frac{D_1-D_2(-1)^l}{D_2(-1)^l}\frac{1}{\eps'}
\left(\frac{Q_1Q_5\omega^4 }{16R^4}\right)^{\frac{l+1}{2}}
\left[\frac{1}{(l+1)!l!}\right].
\ee
Substituting here the value of the ratio (\ref{DRatio}), we finally get
\be
R'=\pi i\frac{1+e^{2\pi i\left(\beta-\alpha+l/2\right)}}
{1-e^{2\pi i\left(\beta-\alpha+l/2\right)}}
\left(\frac{Q_1Q_5\omega^4 }{16R^4}
\right)^{\frac{l+1}{2}}
\left[\frac{1}{(l+1)!l!}\right].
\ee
Making a formal expansion of the denominator we get
\be
R'=\pi i\left(\frac{Q_1Q_5\omega^4 }{16R^4}
\right)^{\frac{l+1}{2}}
\left[\frac{1}{(l+1)!l!}\right]\left\{1+2\sum_{n=1}^\infty
e^{2\pi in\left(\beta-\alpha+l/2\right)}\right\}
\ee
Then the leading contribution to the probability $P_2$ is
\be
S_2=4\pi^2 \left(\frac{Q_1Q_5\omega^4 }{16R^4}
\right)^{{l+1}}
\left[\frac{1}{(l+1)!l!}\right]^2
\ee
We again note that $P_2=S_2^2$, which was anticipated. The probability $P_2$
of going in the throat coincides with absorption coefficient  $a_l$ 
for the unexcited D1-D5
geometry,  which was evaluated in \cite{ms}.

\end{document}